\documentclass[letterpaper,nofootinbib,11pt]{revtex4}

\usepackage{amsfonts}
\usepackage{amsmath}
\usepackage{amssymb}
\usepackage{amsthm}
\usepackage{mathrsfs}
\usepackage{bbm}
\usepackage{bm}
\usepackage{cancel}
\usepackage{graphicx}
\usepackage[usenames,dvipsnames]{color}
\usepackage{palatino}
\usepackage{fancyhdr}
\usepackage{mathrsfs}
\usepackage{wrapfig}

\newcommand{\be}{\begin{equation}}
\newcommand{\ee}{\end{equation}}
\newcommand{\ba}{\begin{array}}
\newcommand{\ea}{\end{array}}
\newcommand{\bea}{\begin{eqnarray}}
\newcommand{\eea}{\end{eqnarray}}
\newcommand{\ssum}{\displaystyle\sum\limits}

\newcommand{\ms}{\mathscr}

\newcommand{\R}{\mathbb{R}}

\newcommand{\fn}{\footnote}
\newcommand{\Rs}{\mathcal{R}}

\newcommand{\st}{\scriptstyle\textrm}
\newcommand{\hsi}{\hbar_{\scriptstyle\textrm{si}}}

\theoremstyle{plain}
\newtheorem{thm}{Theorem}[section]
\newtheorem{corol}[thm]{Corollary}
\newtheorem{lem}[thm]{Lemma}
\newtheorem{propos}[thm]{Proposition}
\theoremstyle{definition}

\theoremstyle{remark}

\usepackage{hyperref}

\hypersetup{
   colorlinks=true,         
   linkcolor=blue,          
   citecolor=red,        
   urlcolor=Violet            
}

\begin{document}

\title{\Large Scale Anomaly as the Origin of Time}

\author{\bf Julian Barbour$^{1,2}$, 
Matteo Lostaglio$^3$\footnote{This work has been submitted in partial fulfillment of the Master Degree in Physics at the University of Pavia.}, {\rm and}
Flavio~Mercati$^{4,5,}$}
\email{barbourj@physics.ox.ac.uk, matteo.lostaglio12@imperial.ac.uk, fmercati@perimeterinstitute.ca}
\affiliation{\it $^1$College Farm, South Newington, Banbury, Oxon, OX15 4JG, UK,\\
$^2$Department of Physics, Oxford University, Denys Wilkinson Building, Keble Road,  
OX1 3RH, UK,\\
$^3$Dep. of Physics, Imperial College London, Prince Consort Rd, London SW7 2BW, UK,\\
$^4$School of Mathematical Sciences, University of Nottingham, NG7 2RD, UK,\\
$^5$Perimeter Institute for Theoretical Physics, 31 Caroline Street North  Waterloo, ON N2L 2Y5, Canada.}


\begin{abstract}

\vspace{12pt}

\begin{center}{\bf  Abstract}\end{center}

We explore the problem of time in quantum gravity in a point-particle analogue model of scale-invariant gravity. If quantized after reduction to true degrees of freedom, it leads to a time-independent Schr\"odinger equation. As with the Wheeler--DeWitt equation, time disappears, and a frozen formalism that gives a static wavefunction on the space of possible shapes of the system is obtained. However, if one follows the Dirac procedure and quantizes by imposing constraints, the potential that ensures scale invariance gives rise to a conformal anomaly, and the scale invariance is broken. A behaviour closely analogous to renormalization-group (RG) flow results. The wavefunction  acquires a dependence on the scale parameter of the RG flow. We interpret this as time evolution and obtain a novel solution of the problem of time in quantum gravity. 
We apply the general procedure to the three-body problem, showing how to fix a natural initial value condition, introducing the notion of complexity. We recover a time-dependent Schr\"odinger equation with a repulsive cosmological force in the `late-time' physics and we analyse the role of the scale invariant Planck constant. We suggest that several mechanisms presented in this model could be exploited in more general contexts. 
%
%
\end{abstract}

\maketitle

\newpage

\tableofcontents

\section{Introduction}

One important difficulty in canonical quantum gravity is the
\emph{problem of time} \cite{kuchar:prob_of_time,Isham:pot_review,Anderson:review_pot}, which arises from the refoliation invariance of general relativity (GR).
The Hamiltonian form of general relativity (GR) has a quadratic constraint at each space point and, at the classical level, \emph{many-fingered-time} evolution. Canonical quantization in the case of a spatially closed universe, leads, at least formally, to the Wheeler--DeWitt equation, which does not contain time. This is the \emph{frozen-formalism} problem. As yet, there is no agreed way in which our manifest experience of a flow of time can be matched to structure in the quantum theory. In addition, there are severe technical problems in implementing the Hamiltonian constraints by quantum operators.

In this paper, we shall not address these technical issues but will suggest that the frozen-formalism problem might be solved simultaneously with a further, less often noted problem -- the failure of GR to be exactly scale invariant. This is best explained in terms of the GR initial-value problem, for which the only robust and effective method of solution is York's based on foliation of spacetime by spacelike hypersurfaces of constant mean (extrinsic) curvature (CMC) \cite{York:cotton_tensor,York:york_method_prl,York:york_method_long}. Represented this way, GR is a theory in which conformal three-geometries interact with a single global variable, the rate at which space expands. This global degree of freedom, a last vestige of Newton's absolute space, is why GR is not \emph{exactly} scale invariant. The failure is mysterious.

It is possible to construct an exactly scale-invariant theory of conformal three-geometries interacting with themselves alone -- essentially GR without the global variable \cite{Anderson:2002ey}. There also exists a closely analogous point-particle model of that theory \cite{barbour:scale_inv_particles}. Its key feature is replacement of the standard Newtonian potential by one homogeneous of degree $-2$, which ensures scale invariance. Newtonian gravitational forces are exactly recovered but with additional attractive forces that ensure scale invariance but are vanishingly small except on cosmological scales.

The observation on which this paper is based is that Dirac-style quantization of the theory with this potential leads to a \emph{conformal anomaly}. This breaks the scale invariance and leads to a renormalization-group (RG) flow governed by a dimensionsless scale parameter. We show that this leads simultaneously to `time evolution' governed by the RG scale parameter. \footnote{The idea of considering RG as the origin of time in Shape Dynamics is due to S. Gryb. See also \cite{SeanFlavioEssay}. } We call this \emph{double emergence} and put it forward as a possible resolution of two fundamental problems: the frozen formalism and the lack of exact scale invariance in gravity. 
\footnote{Its predictions can also be compared with the model explored in \cite{BKMpaper, JulianComplexityPaper} with Newtonian (degree $-1$) potential 
in which the single global degree of freedom that breaks scale invariance is traded classically for an 
internal time, after which the remaining dynamical degrees of freedom are all 
scale invariant. There are in fact two ways to achieve scale invariance, one 
(which we study here) stronger than the other (studied in \cite{BKMpaper, JulianComplexityPaper} ).\label{FootOnTheOtherTwoPapers}} 

In this paper we give a self-contained treatment of the classical particle model with the degree $-2$ potential of ref. \cite{barbour:scale_inv_particles} and its quantization with double emergence of the RG scale and an effective time through breaking of the scale invariance. We also treat the three-body problem, which gives us an opportunity to introduce what appears to be an interesting distinguished initial condition `at the origin of RG time'. We also explore a further novel feature of the quantization: it has \emph{dimensionless} Planck constant.

A word about concepts. In the Newtonian three-body problem all $3\times 3$ particle coordinates are classical and quantum observables. But if the three particles are the entire universe, the position, orientation, and size of the triangle they form cease to have any observable meaning. There is also no external clock to measure time. Only the possible shapes of the triangle remain as observables. The collection of all possible shapes is \emph{shape space} $\mathcal{S}$. It is our fundamental concept. We insist that only functions defined on $\mathcal{S}$ are classical or quantum observables. Position, orientation, and size are all gauge. But we shall see that relative size, \emph{without becoming an observable}, can play the role of an independent variable like time.

\section{The Analogue Model}

General relativity in the ADM formulation is a reparametrization-invariant theory, so its
Hamiltonian vanishes on the solutions of the equations of motion. The particle model of \cite{barbour:scale_inv_particles} has a similar dynamical structure to GR but on a different phase space: $\R^{6N}$, the space of the Cartesian coordinates of $N$ point particles ${\bf r}^J=({r_1}^J,{r_2}^J,{r_3}^J)$, or, interchangeably, ${\bf r}^J=({r_x}^J,{r_y}^J,{r_z}^J)$,  together with their conjugate momenta ${\bf p}^J=({p_1}^J,{p_2}^J,{p_3}^J)=({p_x}^J,{p_y}^J,{p_z}^J)$, where $J=1,...,N$. The reduced phase space is defined by the algebra of constraints
\be
H=\ssum_{ J=1}^{ N} \frac{ {\bf p}^{ J}\cdot {\bf p}^{J}}{2\;m_{J}}+V=0 \; ,
\ee
\noindent which implements reparametrization invariance, and
\be
{\bf P}=\ssum_{J=1}^{\rm N}{ \bf p}^{J} = 0 \; , \qquad {\bf L}= \ssum_{J=1}^{\rm N} { \bf r}^{J} \times {\bf p}^{J} = 0 \; ,
\ee
\noindent which implement translational and rotational invariance \cite{barbourbertotti:mach}. If $V$ is a function of the $N(N-1)/2$ separations $||{ \bf r}^J - { \bf r}^I||$ only, the algebra of constraints closes:
\be
\{L_i,L_j\}= \epsilon_{ijk}L_k \; , \quad \{L_i,P_j\}=  \epsilon_{ijk}P_k \; , \quad \{P_i,P_j\}= 0 \; ,
\ee
\be
\{H,P_i\}=\ssum_{J=1}^{N} \frac{\partial V}{\partial r_i^J} = 0 \; , \qquad \{H,L_i\}= \ssum_{J=1}^{N} \epsilon_{ijk} r_j^J \frac{\partial V}{\partial r_i^J} = 0 \; ,
\ee
\noindent where we used Einstein's summation convention and $\{\cdot,\cdot\}$ denotes the Poisson bracket
\be
\{r_i^I,p_j^J\}= \delta_{ij}\delta^{IJ} \; .
\ee
 This defines a theory of the evolution of \emph{relative configurations} that can still differ in size.
To implement scale invariance and define a theory on shape space, we add the constraint
\be\label{DilatationConstraint}
D = \ssum_{J=1}^N {\bf r}^J \cdot {\bf p}^J \; .
\ee
Following \cite{barbour:scale_inv_particles} we call $D$ the \emph{dilational momentum}. It has the same dimensions as the angular momentum $\textbf{L}$ and measures expansion of the system just as $\textbf{L}$ measures rotation. A scale-invariant theory clearly cannot allow physical expansion, for that requires an absolute scale. Vanishing of (\ref{DilatationConstraint}) is a necessary condition for scale invariance.
But it is not sufficient: (\ref{DilatationConstraint}) imposes further constraints on $V$.
Indeed, the Poisson algebra now also includes
\be\label{ConformalConstraintAlgebra}
\{D , P_j\} = - P_j\; , \qquad \{ D, L_j \} = 0 \; , \qquad \{ H , D \} = - 2 \, H +
 \ssum_{J=1}^{N}  r_j^J \frac{\partial V}{\partial r_j^J} + 2 \, V \;,
\ee
and it closes if $\sum_{J=1}^{N}  r_j^J \frac{\partial V}{\partial r_j^J}  + 2 \, V = 0$.
By Euler's theorem this requires $V$ to be homogeneous of degree $-2$ 
in the coordinates $r^I_j$. The potential with this property chosen in \cite{barbour:scale_inv_particles} is
\be
V=\frac{V_{\rm New}}{R} \,, \qquad V_{\rm New} = - \sum_{I<J} \frac{m_I \, m_J}{||{\bf r}^I - {\bf r}^J||} \; , \qquad R^2=\frac{1}{M^2}\ssum_{I<J=1}^{N} m_J \, m_I \, ||{\bf r}^I - {\bf r}^J||^2 \;,
\ee
where $M$ is the total mass and $MR^2$ is the centre-of-mass moment of inertia. The classical mechanics of this model reproduces Newtonian gravity to high accuracy in small enough subsystems but with a long-range attractive cosmological force that prevents the universe from expanding \cite{barbour:scale_inv_particles}. Indeed $R$ is a constant of motion. 

In this paper we will study the quantum mechanics of this model. It turns out to be rich in interesting features.
We first discuss an old issue in the quantization of constrained systems. 

\section{Quantization: Before or After Reduction?}

There are at least two ways to build a relational quantum theory. At first glance it seems most natural to get rid of the coordinate redundancies immediately and to quantize directly the true degrees of freedom. The other possibility is to impose the constraints quantum mechanically: this is Dirac quantization \cite{dirac:lectures}. The two procedures are in general inequivalent, so which is the correct one?

 The motivation for quantization after reduction is that the physical degrees of
freedom live in the reduced phase space and therefore all the remaining ones are redundant (gauge) and unphysical. However, there are severe mathematical difficulties, generally insuperable, associated with working with true degrees of freedom, and for this reason alone the redundant degrees are usually retained.

We have four reasons not related to mere computational convenience to do the same:

1. There are several examples in physics of \emph{anomalies}, in which a classical symmetry
is not preserved by the quantum theory and the consequences have been confirmed experimentally. 
Quantization is by no means a unique procedure, as already operator ordering issues show,
and a single classical theory usually corresponds to the classical limit of several quantum
theories.
Ultimately the observations tell us which quantization procedure provides the theory that fits them better.

2. In gauge theories the interactions are local when expressed redundantly but most awkwardly nonlocal in true degrees of freedom.

3. The laws of transformation from one set of redundant coordinates to another are very simple. In fact, Cartesian coordinates, like many other examples of redundant coordinates associated with an underlying Lie group, are crucial in suggesting the form of the kinetic energy. The point is that there is no unique prescription for choosing true degrees of freedom and, accordingly, no indication of how much kinetic energy should be associated with their rates of change. In contrast, the Cartesian coordinates are distinguished and \emph{democratic}: weighted with the respective particle masses, they provide an obvious measure of the kinetic energy that should go with each. The residual ambiguity that arises from time-dependent group transformations between Cartesian frames is eliminated by means of constraints. Moreover, this approach selects a preferred form of these constraints, whereas 
in general one could get classically equivalent but quantum-mechanically 
different theories simply by multiplying the original constraints by arbitrary 
regular functions. This ambiguity is resolved if one works with group-
theoretically defined redundant coordinates, which are distinguished and select 
the usual form of the Hamiltonian constraint.

4. The reduction before quantization approach also runs into the problem of recovering a meaningful notion of evolution in reparametrization-invariant theories like GR.  





\section{Quantization of the Analogue Model}

\subsection{Eliminating the momentum and angular momentum constraints}

It is easy to show that the two quantizations are equivalent for the momentum constraint $\bm P=0$, which can be conveniently
solved by moving to mass-weighted \textit{Jacobi coordinates} $\{ {\bm \rho}^K\}$, $K=1,...,N-1$. 
These are a relational $N$-body generalization of the reduced-mass coordinates
used to decouple the centre-of-mass motion in the two-body problem.
There are many choices of Jacobi coordinates for one and the same $N$-body system. We refer to the Appendix \ref{JacobiCoordAppendix} for details.

Suppose we have chosen Jacobi coordinates $\{{\bm \rho^J} \}$. We denote by $\{ \bm \pi^J \}$, $\bm \pi^J = ({\pi_1}^J,{\pi_2}^J,{\pi_3}^J)$, the momenta conjugate to $\{\bm \rho^J \}$.
In the new coordinates, the moment of inertia is mapped into the \textit{barycentric moment of inertia} \cite{LittlejohnReinsch}:
\be
R^2 = \ssum_{J=1}^{N-1} {\bm \rho}^J \cdot {\bm \rho}^J \; .
\ee
The new coordinates automatically implement the constraint $\textbf{P}=0$, so we just need to consider the constraints $\textbf{L}=0$ and $D=0$. The dilatational momentum constraint (\ref{DilatationConstraint}) becomes
\be
D= \ssum_{J=1}^{N-1} {\bm \rho}^J \cdot {\bm \pi}^J = 0 \; ,
\ee
and the Hamiltonian and angular momentum constraints are \cite{Anderson:review_pot}
\be \label{ClassHamConstr}
H = \ssum_{J=1}^{N-1} \frac{{\bm \pi}^J \cdot {\bm \pi}^J}{2} + \frac{V_{\rm New}}{R} = 0 \;, \qquad 
 {\bf L}= \ssum_{J=1}^{\rm N} { \bm \rho}^{J} \times {\bm \pi}^{J} = 0 \; .
\ee

The constraint $\textbf{L}=0$ can be solved by a gauge fixing (see Appendix IX B).\footnote{The $SO(3)$ gauge symmetry of $N$-body dynamics is probably responsible for a lot of interesting
features of this model, in particular at the quantum level, as shown by the wealth of results obtained
by Littlejohn and Reinsch \cite{LittlejohnReinsch}, who took seriously the fibre-bundle structure of the configuration space.}
However, our main focus is scale invariance. 
From now on, therefore, we will keep the rotational gauge unfixed and solve the constraint $\textbf{L}=0$ by Dirac quantization. An exception
will be the three-body problem, which is always \emph{planar} when $\textbf{L} = 0$ and has a simple algebraic phase-space reduction.

\subsection{Quantization of the Hamiltonian constraint}
The Poisson brackets between $ {\bm \rho}^J$ and ${\bm \pi}^J$ are
\be
\{ \rho^I_i, \pi^J_j \} = \delta_{ij} \, \delta^{IJ} \,.
\ee
We quantize these relations by promoting $\rho^I_i$ and
$\pi^J_j$ to operators $\hat \rho^I_i$, $\hat \pi^J_j$ that act on the Hilbert space $\ms{L}^2(\R^{3N-3})$ and 
close the Heisenberg algebra 
\be
[ \hat \rho^I_i, \hat \pi^J_j ] = i \; \hbar_{\st{si}} \; \delta_{ij} \, \delta^{IJ} \,.
\ee
A caveat: $\hbar_{\st{si}}$ \emph{is not}~the familiar Planck constant, but
a dimensionless number ($\hat \rho^I_i$ has the dimensions of a length, while $\hat \pi^J_j $
is a length$^{-1}$). We will see in Sec.~\ref{latetime} how may be related to the usual dimensionful Planck constant.

The quantum Hamiltonian constraint (\ref{ClassHamConstr}) is
\be
\hat H \; \psi = \left(- \frac{1}{2} \hsi^2 \, \Delta_{3N-3} +\frac{V_{\rm New}}{R}  \right) \psi = 0 \,,
\ee
where $\Delta_{3N-3}$ is the Laplacian in $3N-3$-dimensional Euclidean space.

In polar coordinates
$(\rho_1^1,\rho_2^1,\rho_3^1 , \dots ,\rho_1^{N-1},\rho_2^{N-1},\rho_3^{N-1})  \to (R, \sigma)$, $R \in \R^+$, $\sigma \in S^{3N-4}$\;, 
\be
\label{hamiltonianconstraint}
\hat H \; \psi(R,\sigma) = \frac{1}{2} \, \left[ - \hsi^2 \, \frac{\partial^2}{\partial R^2}  - \hsi^2 \,\frac{3N-4}{R} \frac{\partial}{\partial R} + \frac{1}{R^2} \left( - \hsi^2 \, \Delta_{S^{3N-4}} + 2 \, V_{\st{shape}} \right) \right] \psi(R,\sigma) = 0 \,,
\ee
where $ \Delta_{S^{3N-4}} $ is the spherical Laplacian, and we call $V_{\st{shape}} =  R \,V_{\st{New}} $ the \emph{shape potential} \cite{barbour:scale_inv_particles,barbourbertotti:mach}.\footnote{Saari \cite{SaariBook} calls $V_{\st{shape}}$ the \emph{configuration measure.}} Its very interesting properties are investigated in the companion paper \cite{JulianComplexityPaper}. 

The reparametrization invariance of a relational theory like ours leads to a time-independent Schr\"odinger equation \eqref{hamiltonianconstraint}, the analogue in particle systems of the timeless Wheeler--DeWitt equation in geometrodynamics.

\subsection{Separation of the problem}

The potential $V=V_{\st{New}}/R$ allows us to separate the variables in Eq.~\eqref{hamiltonianconstraint}\,\footnote{We shall show that the functions $\varphi_n$ are
countable, so we adopt this notation already.}
\be
\label{separation}
\psi(R,q) = \ssum_n \; \xi_n(R) \; \varphi_n (\sigma) \;.
\ee
Substitution of this ansatz in \eqref{hamiltonianconstraint} gives, for each $n$,
the eigenvalue equation
\be \label{scaleinvariantproblem}
\hat H_{\st{shape}} \, \varphi_n = \left( - \hsi^2 \,\Delta_{S^{3N-4}} + 2\;V_{\st{shape}} \right) \varphi_n =  \lambda_n \; \varphi_n \;;
\ee
this is a ``scale-invariant equation''.\footnote{\label{capsule} One of the authors has proposed, in \cite{barbour:timelessness2}, that a notion of time might be
hidden in the frozen wavefunction resulting from the reduction before quantization approach.
This might be the case if the wavefunction of the universe is concentrated on `time capsules', \emph{i.e.}, configurations which suggest a history. In the reduction-before-quantization approach, the only allowed solution of (\ref{scaleinvariantproblem}) is one with $\lambda_n=0$, corresponding to zero total energy and a frozen wavefunction. 
Rewriting (\ref{scaleinvariantproblem}) in the form $\hat H_{\st{shape}} \, \varphi_n = - \Delta_{S^{3N-4}}\varphi_n + (2/ \hsi^2) V_{\st{shape}} \varphi_n =  (\lambda_n /\hsi^2)\; \varphi_n $, we see that $2/\hsi^2$ plays the role of a dimensionless coupling constant. It might perhaps be possible to adjust its value to ensure the existence of the zero eigenvalue. This would be a novel first-principles derivation of the strength of quantum effects.} 
The operator $\hat H_{\st{shape}}$, which acts only on the shape variables, is self-adjoint on the dense domain $H^2(S^{3N-4})$ (Sobolev space of $\ms L^2$ functions whose derivatives up to the second order are $\ms L^2$). By the Weyl and Rellich--Kato theorems \cite{SimonsReed} and the compactness of $S^{3N-4}$ it can be proved it has a bounded from below and discrete spectrum.

We still have to impose the constraints $\hat L_j \; \psi =0$ and $\hat D \; \psi = 0$.
The former acts only on the scale-invariant wavefunctions because it commutes with both $R$ and $\partial / \partial R$. $\hat L_j$ commutes also with $\hat H_{\st{shape}}$, and therefore the remaining constraint 
$\hat L_j \; \varphi_n = 0$ just acts as a selection rule on the scale-invariant eigenfunctions $\varphi_n$. 
Applying this selection rule gives solutions defined on \emph{shape space}, that
is  $\R^{3N}/ \text{\it Similarity}(\R^3)$.

For each shape eigenvalue $\lambda_n$, we get a radial equation on $\ms{L}^2(\R^+)$:
\be 
 \left( - \hsi^2 \frac{\partial^2 }{\partial R^2} - \hsi^2 \frac{3N-4}{R} \frac{\partial}{\partial R} + \frac{\lambda_n}{R^2}\right) \xi_n = 0 \;,
\ee
which can be nicely rewritten by means of the change of variables
$ u_n(R)=(\mu R)^{(3N-4)/2}\xi_n(R) $, where $\mu>0$ is an arbitrary scale with dimensions length$^{-1}$, introduced to keep $u_n$ dimensionless:
\be
\label{radial}
\hat H^n_{\rm rad} \; u_n = - \hsi^2 \frac{\partial^2 u_n}{\partial R^2} + \frac{g_n}{R^2}u_n = 0 \;,  \qquad 
g_n = \lambda_n - \frac{1}{4} (3N-4)(3N-6) \; \hsi^2\;.
\ee

\subsection{The anomaly}

The anomalous behaviour of a $1/R^2$ potential like the one in the radial equations Eq.~\eqref{radial} has been well-known since the pioneering work of Case \cite{CSP} in the 1950s. Initially dismissed as non-physical, its importance has only relatively recently been recognized in very different areas, from molecular physics to black holes  (see \emph{e.g.} \cite{CMB} and the references therein). 
In particular there has been a surge of interest after the experimental confirmation \cite{IOF,KEE} of the Efimov effect \cite{EEE}, which is now better understood as a consequence
of this quantum anomaly. 

The novelty of our approach is that we found, starting from relational first principles, similar equations describing a cosmological setting. A difference is that each radial equation is coupled with scale-invariant degrees of freedom and ultimately, through the anomaly, will determine their evolution in an emergent time. Let us develop the argument.

We can see that a $1/R^2$ potential can generate an anomaly in this way: consider the
quantization of the dilatation constraint \eqref{DilatationConstraint}. There is a
mild ordering ambiguity. Without loss of generality we will take the Weyl ordering
\be
\hat D =  \frac 1 2 \ssum_{J,i} \left( \hat \rho^J_i \,\hat \pi^J_i +
\hat \pi^J_i \hat \rho^J_i  \right) \equiv -\frac{i}{2} \left( R \frac{\partial}{\partial R} + \frac{\partial}{\partial R}R \right)  \;.
\ee
From Eq. \eqref{ConformalConstraintAlgebra} and the canonical quantization we get
\be
\label{c3HDcom}
[\hat H , \hat D] \psi = - 2 \; \hat H \, \psi + i \, \hsi \left( 2 \; V  + \ssum_{J=1}^{N-1} \rho^J_i \frac{\partial V}{\partial \rho^J_i} \right) \psi := -2\hat H\, \psi + i \, \hsi \; \hat A \; \psi \;.
\ee
Classically, the quantity $\hat A$ vanishes for a $1/R^2$ potential, due to Euler's 
theorem, but quantum mechanically this theorem breaks down, as we will see. The anomaly can be rewritten:
\be
\label{aanomaly}
\hat A = 2V +  \underline{\rho} \cdot \nabla(V\cdot) = 2\frac{R^{3N-5}}{2} \nabla \cdot \left( \frac{\underline{\rho} V}{R^{3N-5}}\right) \; ,
\ee
where $\underline{\rho}~=~(\rho_1^1,\dots, \rho_i^J,\dots,\rho_3^{N-1})$ and $\nabla=(\partial/\partial \rho_1^1,\dots,\partial/\partial \rho_i^J,\dots,\partial/\partial \rho_3^{N-1})$.
%
%
%
%
\be
\nonumber
\hat A = 2\frac{R^{3N-5}}{2} \nabla \cdot \left( \frac{\underline{\rho} V_{\st{shape}}}{R^{3N-3}}\right)= 2\frac{R^{3N-5}}{2}\nabla \cdot \left(\frac{\underline{\rho}}{R^{3N-3}}\right)V_{\st{shape}}+ \frac{1}{R^2}\underline{\rho} \cdot \nabla V_{\st{shape}} \; .
\ee

\noindent Defining $\Omega_{3N-4}$ as the surface of a $3N-4$ dimensional sphere, we have
\be
\label{distributional}
\nabla \cdot \left(\frac{\underline{\rho}}{R^{3N-3}}\right)= \Omega_{3N-4} \delta^{3N-3}(\rho),
\qquad
\hat A= R^{3N-5} \Omega_{3N-4}  \delta^{3N-3}(\underline{\rho}) V_{\st{shape}} + \frac{1}{R^2}\underline{\rho} \cdot \nabla V_{si} \; .
\ee
Equation \eqref{distributional} is exactly where the classical Euler theorem departs from the quantum one. 
Now $\delta^{3N-3}(\underline{\rho})=\frac{1}{\Omega_{3N-4}R^{3N-4}}\delta(R)$, and because $V_{\st{shape}}$ does not depend on $R$ Euler's theorem holds for it. This implies
\be
\hat A=V_{\st{shape}}R^{-1}\delta(R) \; .
\ee
The distribution $A$ is called an \emph{anomaly}. Our algebra does not close even if the potential is homogeneous of degree $-2$,
\be
[\hat H,\hat D] = -2\hat H + i\hbar V_{\st{shape}}R^{-1} \delta(R) \; ,
\ee
\textit{unless} we ask for particular boundary conditions at the origin. We can interpret this result as a failure of Euler's theorem at the quantum level \cite{AAC,CMB}. In the next section we will see that the closure of the algebra is incompatible with self-adjointness of the radial Hamiltonian. We should emphasize that self-adjointness is a working hypothesis that we retain in a theory of the whole universe, for which it is not \emph{a priori} clear whether all the basic postulates of quantum theory, established in subsystems of the universe, are preserved. \emph{Caveat lector.}

\subsection{Self-adjoint extensions}

The radial equations \eqref{radial} on the dense domain 
\be
\ms{D}(H^n_{rad})=\{u, \; u' \; \textrm{absolutely continuous and } \; u(0)=u'(0)=0 \} \; ,
\ee
have been studied extensively in the literature \cite{CMB}. The deficiency indices depend on the energy regime. They are
\begin{enumerate}
\item $(0,0)$ ~�if ~ $g_n/\hsi^2 > 3/4$, ~so that the radial Hamiltonian is self-adjoint on $\ms{L}^2(\R^+)$.
\item $(1,1)$ ~ if ~ $-1/4 <g_n/\hsi^2 <3/4$, ~so that there is a $U(1)$ group of self-adjoint extensions, labelled by the index $\theta$ . Scale invariance is generally broken, but there are two conformal fixed points of the renormalization group (RG) flow. We will call this the \textit{weak coupling regime}.
\item $(1,1)$ ~ if ~�$g_n/\hsi^2 < -1/4$, ~ but each element of the $U(1)$ group of self-adjoint extensions breaks scale invariance, \emph{i.e.}, there are no conformal fixed points in the RG flow. A limit cycle behaviour emerges. We will call this the \textit{strong coupling regime}. 
\end{enumerate}
From the fundamental characterization of self-adjoint extensions \cite{SimonsReed}, the domain of $\hat H^n_{\theta}$ ~ is the direct sum of ~ $\ms{D}(H^n_{rad})$ ~ and the span of
\be
\label{wtheta}
w_{\theta}(\Rs)= \sqrt{\Rs}\left[ K_{\Xi_n}(e^{i\pi/4}\Rs) + e^{-i\theta}K_{\Xi_n}(e^{-i\pi/4}\Rs)\right] \; , \qquad \Xi_n=\sqrt{g_n/h^2_{si} +1/4} \;,
\ee
consisting of a linear combination of the solutions of $\hat H^n_{rad} u = \pm \, i \, \mu \, u$; we need $\mu$  for dimensional consistency; $\mathcal{R}= \mu R/ \hsi$, and $\Xi_n$ is the order of the Bessel function $K_{\Xi_n}$. The self-adjoint extensions break scale invariance, generating `bound states' \cite{CMB} with nonzero eigenvalues. With $\hat H^n_{\theta}$ denoting the $\theta$-extension of the radial operator $\hat H^n_{\st{rad}}$\,, the eigenstates are given by the equation
\be
\label{c4radial}
\hat H^n_{\theta}u = -\kappa^2 u \qquad \Longrightarrow \qquad
-\frac{\partial^2 u}{\partial \Rs ^2} + \frac{g_n}{\hsi^2}\frac{u}{\Rs^2} = - \frac{\kappa^2}{\mu^2}u \; .
\ee
The substitution $u(\Rs)=\sqrt{\kappa \Rs/\mu }f(\Rs)$ leads us to a modified Bessel equation for $f$; the two independent solutions are by definition
\be
u(\Rs)=\sqrt{\frac{\kappa}{\mu}  \Rs}\; K_{\Xi_n}\left(\frac{\kappa}{\mu} \Rs \right)\; , \; \; \; \; \; \; v(\Rs) = \sqrt{\frac{\kappa}{\mu} \Rs} \; I_{\Xi_n}\left(\frac{\kappa}{\mu} \Rs \right) \; ,
\ee
but, for both real and imaginary values of $\Xi_n$, only the first one is a solution in $\ms{L}^2(\R)$.

To choose a self-adjoint extension amounts to imposing a boundary condition at the origin in the previous problem, so we now want to understand which eigenfunctions $\sqrt{\kappa \Rs/\mu}\;K_{\Xi_n}\left(\frac{\kappa}{\mu} \Rs \right)$, corresponding to eigenvalues $-\kappa^2$, are compatible with the boundary condition
\be
\label{c4boundarycond}
\lim_{\Rs \rightarrow 0} u(\Rs) = \lim_{\Rs \rightarrow 0} w_{\theta}(\Rs) \;.
\ee
This condition gives a relation between $\kappa$ and $\theta$. Expanding $ w_{\theta}$ and $u$ near the origin, we get
\be
\label{kvstheta}
\frac{\kappa_n^2}{\mu^2}=\exp \left[\frac{1}{\Xi_n} \log \left(\frac{\cos(\theta/2 - \pi \, \Xi_n/4)}{\cos(\theta/2 + \pi \, \Xi_n/4)} \right)\right] \;.
\ee

\begin{figure}[b]
\begin{minipage}[h!]{0.46\textwidth}
\centering
\includegraphics[width=\textwidth]{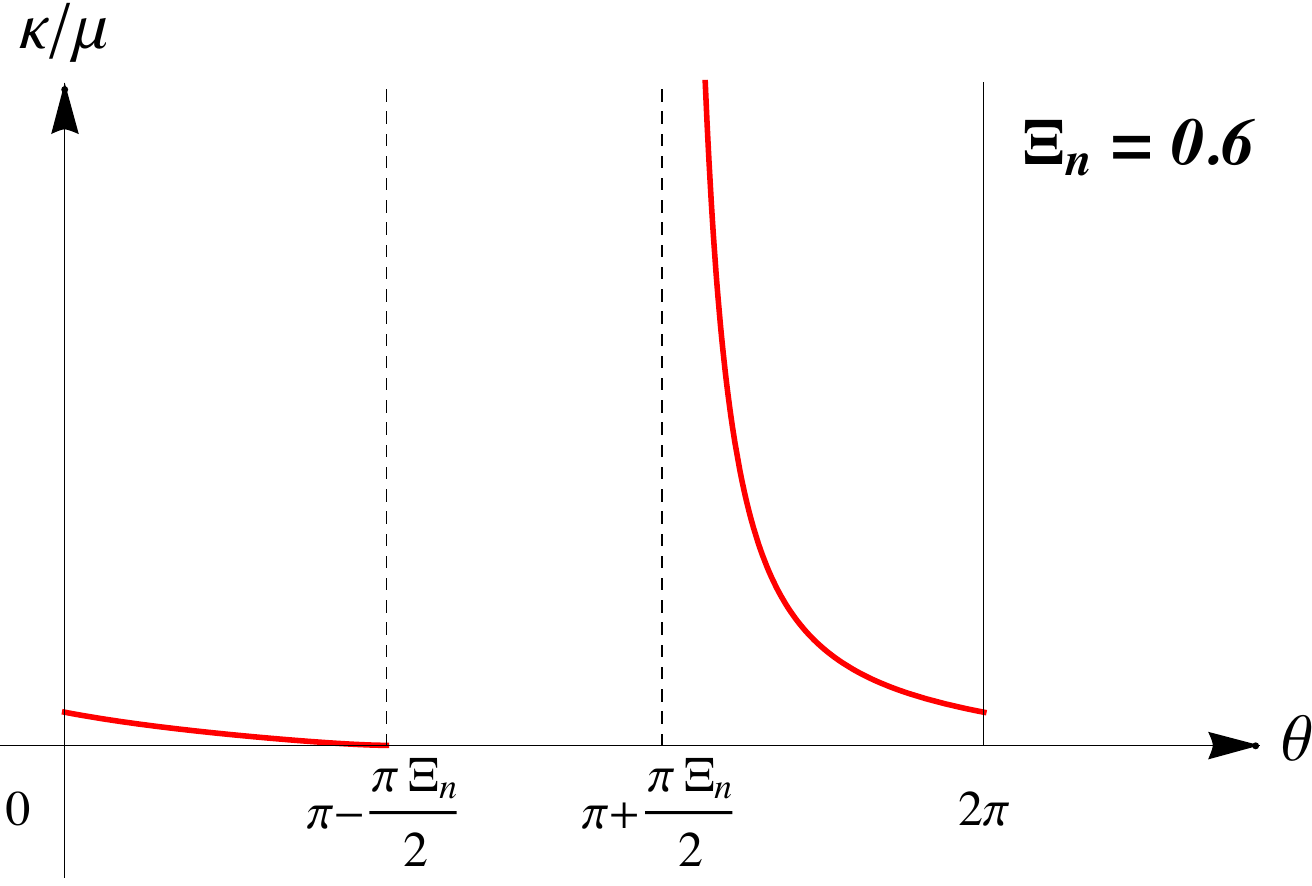}
\end{minipage}
\hspace{1cm}
\begin{minipage}[h!]{0.46\textwidth}
\centering
\includegraphics[width=\textwidth]{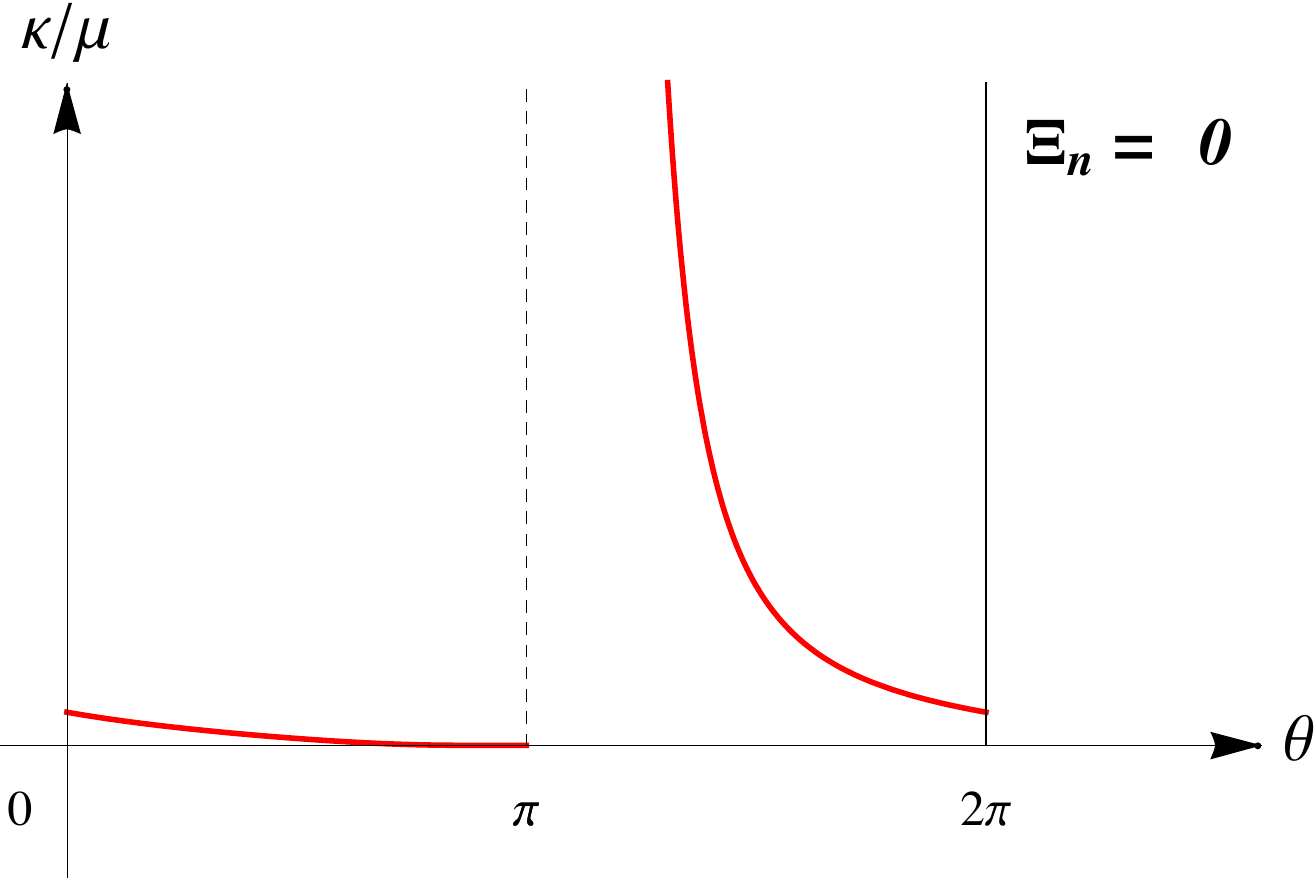}
\end{minipage}
\hspace{10mm}
\caption{$\kappa$ vs $\theta$ in the weak regime. The two preferred choices of $\theta$ correspond to $\kappa=0$ and $\kappa =\infty$. The boundary between the weak and strong regimes, where the two distinguished self-adjoint extensions merge, is at $\Xi_n=0$.}
\label{cylinderweak}
\end{figure}

The weak and strong coupling regimes have qualitatively different behaviours. 
In the weak coupling regime
$\Xi_n \in [0,1]$ is real and Eq. \eqref{kvstheta} is one-to-one, so each self-adjoint extension selects one and only one state of the discrete spectrum. The function $\kappa_n (\theta)$ is defined only in the domain  $\theta~\in~[0,\pi - \pi \, \Xi_n/2] \bigcup (\pi + \pi \, \Xi_n / 2, 2 \pi]$, where it is bijective (see Fig. \ref{cylinderweak}, left-hand side). 
The two boundary points, $\pi - \pi \, \Xi_n/2$ and $\pi + \pi \, \Xi_n / 2$, corresponding respectively to the values $\kappa = 0$ and $\kappa = \infty$,
merge at the transition $\Xi_n \to 0$ between the weak and strong coupling regimes (Fig. \ref{cylinderweak}, right-hand side) .
This is relevant for the RG interpretation of the problem, as we will see later.

\begin{figure}[t]
\centering
\includegraphics[width=0.5\textwidth]{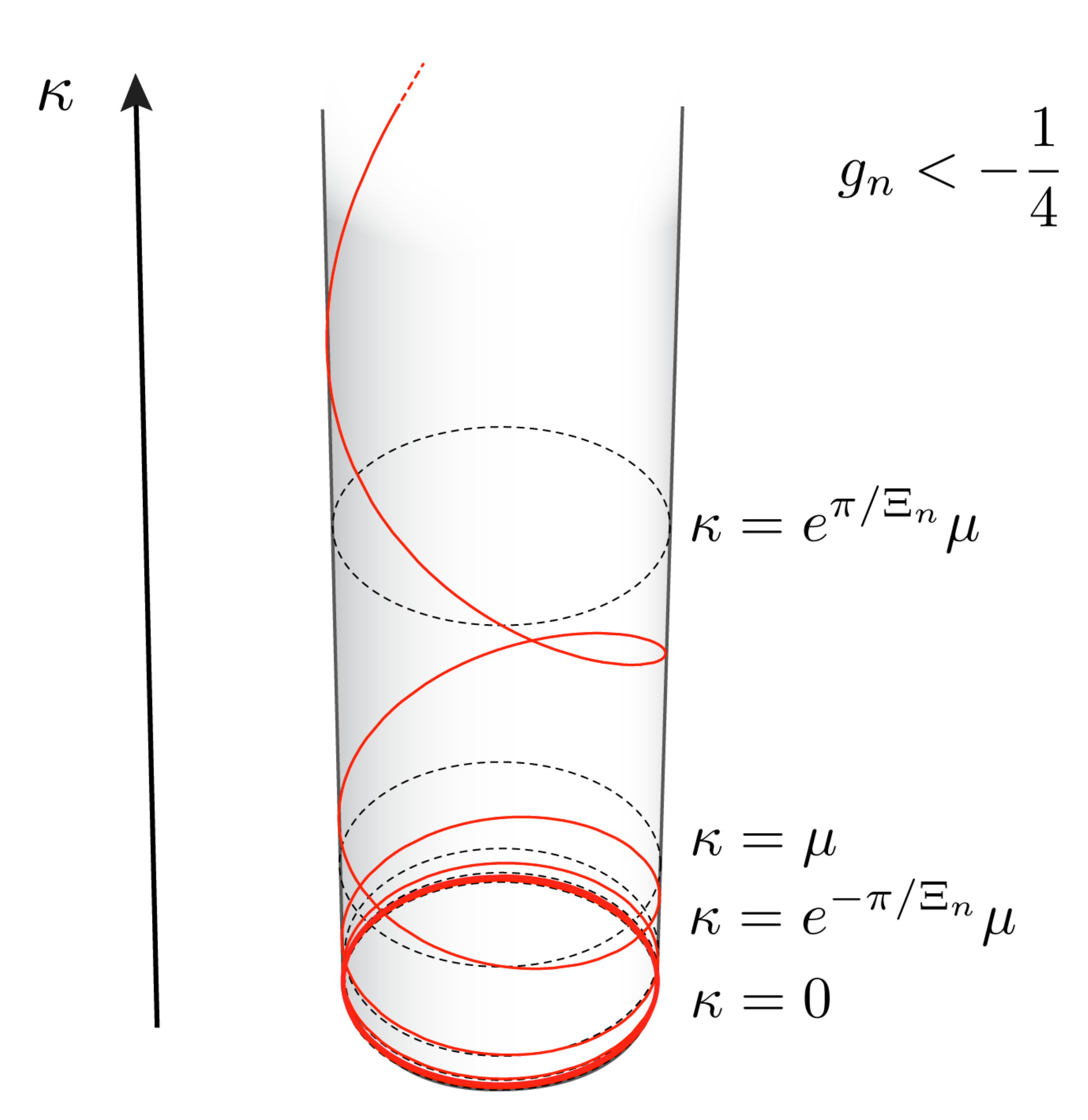}
\caption{$\kappa/\mu$ vs $\theta$ in a cylindrical representation in the strong regime. At each $\theta$ there is a conformal tower of states in a geometric sequence: the ratio between two neighbouring levels is a constant fixed by $\Xi_n$. Continuous variation of $\theta$ over a $2 \pi$ range, leads from one level of the tower to the next.}
\label{cylinderstrong}
\end{figure}

In the strong attractive regime $\Xi_n$ is imaginary, and this causes an abrupt change. The function $\kappa(\theta)$ is defined on the whole range $\theta \in [0,2\pi)$, but is multivalued:
in this case Eq. \eqref{kvstheta} can be rewritten as
\be
\frac{\kappa_{nm}^2}{\mu^2} = \exp \left[ -\frac{\theta}{|\Xi_n|} + \frac{2}{|\Xi_n|} \arctan \left(\frac{\sin \theta}{\cos{\theta} + e^{\pi |\Xi_n|/2}}\right) \pm \frac{2 \pi \, m}{|\Xi_n|}\right] \;, \qquad m \in \mathbb N \;.
\ee


Each $\theta$ selects an entire discrete spectrum made of a tower of states with an accumulation point at zero. These are all the values of the energy compatible with the boundary condition fixed by the chosen $\theta$. The motion of each level as $\theta$ is allowed to vary leads to a limit cycle behaviour, as explained in Fig. \ref{cylinderstrong}.

If we imagine the line twisting around the cylinder as a spring, a change in $\Xi_n$ corresponds to a change in its elongation. In fact the ratio between one level $E^n_m$ and the one immediately below, $E^n_{m-1}$, is equal to a constant that depends on $\Xi_n$: $E^n_m / E^n_{m-1} = e^{\pi /|\Xi_n|}$.
This is known as the Efimov effect \cite{EEE}; it is characterized by the presence of a `conformal tower' of states in a geometric sequence $\{...,(\frac{\kappa_n}{\mu}) e^{-2\pi/|\Xi_n|},(\frac{\kappa_n}{\mu}) e^{-\pi/|\Xi_n|},(\frac{\kappa_n}{\mu}), (\frac{\kappa_n}{\mu}) e^{\pi/|\Xi_n|}, (\frac{\kappa_n}{\mu}) e^{2\pi/|\Xi_n|},... \}$.




\section{Renormalization Group Time}

\subsection{The Callan-Symanzik equation }

The anomaly broke the constraint $\hat D \, \psi=0$, which does not need to be
satisfied by the wavefunction anymore. But there is a domino effect involving the constraint $\hat H \, \psi= 0$, which turns out to be incompatible with the requirement of self-adjointness of the radial Hamiltonian.\fn{This is just a consequence of the fact that the eigenvalue zero in the strong regime is an accumulation point in the spectrum of every self-adjoint extension of each radial Hamiltonian without being an eigenvalue associated with a proper eigenfunction.} We will shortly argue that this generates an RG flow.

The minimal ansatz that makes the theory consistent is to relax the Hamiltonian constraint by introducing an
energy eigenvalue $-\kappa^2$,
\be
\hat H \, \psi = - \kappa^2 \, \psi \;, \label{NewHamiltonianConstraint}
\ee
which is not in conflict with the other constraints $\hat {\bf P} \, \psi = 0$ and $\hat {\bf L } \, \psi = 0$. Support for this minimal ansatz is that it maintains the reparametrization invariance of the classical theory, though no longer with energy eigenvalue $E=0$. 
Moreover from Eq. \eqref{kvstheta} this requirement fixes one and only one self-adjoint extension for each radial equation \eqref{radial} at the subtraction point. They are the extensions $\bar{\theta}_n$ such that
\be
\kappa_{nm}(\bar{\theta}_n) = -\kappa^2 \; \; \forall \, n \;.\fn{The label $m$ does not affect the choice of $\theta_n$ in the strong regime because it corresponds to a complete turn around the cylinder (\ref{cylinderstrong}). Moreover in the weak regime the label $m$ is not present at all.}
\ee
In other words, the single ansatz \eqref{NewHamiltonianConstraint} not only makes the theory self-consistent but also fixes all the mathematical redundancies that arise from the theory of self-adjoint extensions. 

Let us now study how the notion of time emerges in our model as an RG flow.
The wavefunction $\psi$ splits into a sum in which
the coefficients $u_n$ will be interpreted as couplings for the scale-invariant
eigenfunctions $\varphi_n$,
\be
\label{psi}
\psi(R,\sigma) = \sum_n \,  \, u_n(R) R^{(4-3N)/2} \, \varphi_n (\sigma) \;, \qquad \sigma \in S^{3N-4} \;;
\ee
the couplings $u_n$ flow differently under the RG flow, as detailed in the following subsection, in a way that is controlled solely by the
scale-invariant energies $g_n$ with which each is associated. In other words, each scale-invariant eigenfunction determines its own RG evolution. Moreover, from \eqref{psi} we see that an evolution of the $u_n$'s amounts to a redistribution of the relative amplitudes of the scale-invariant eigenfunctions. This redistribution looks very similar to an \textit{evolution with respect to a `time' that is the scale parameter of the RG flow}. \footnote{An advantage of this proposal is that the usual internal time scenarios generically have great difficulty in meeting the requirement of monotonicity whereas in our framework time, being associated to RG flow, is monotonic by definition.}

This might appear rather formal, but the language of effective field theories provides us with a physical interpretation of what is going on. The Hamiltonian constraint implies Eq. (\ref{c4radial}) for the $u_n$'s.
This equation can be recast in an interesting form if we re-express it as
an equation for the \emph{logarithmic derivative} of $u_n$:
\be
\gamma_n (R) = \frac{R \, \frac{\partial u_n(R)}{\partial R} }{u_n(R)} \;, \qquad u_n(R) \propto e^{\int_0^{\log R} \,\gamma_n(R') \, d \log R'} \;.
\ee
The radial equation for $\gamma_n$ becomes
\be \label{RGflowEquations}
\frac{\partial \gamma_n}{\partial \log R} = -  \gamma_n^2 + \gamma_n + \frac{g_n}{\hsi^2} +R^2 \, \kappa^2 \;.
\ee
This last equation has the form of a Callan-Symanzik equation 
for a coupling 
$\gamma_n$ with beta-function
\be \label{BetaFunction}
\beta(\gamma_n , R) = \frac{\partial \gamma_n}{\partial \log R} = - \left( \gamma_n - 1/2 - \sqrt{ \Xi_n^2+ R^2 \kappa^2 }\right) \left( \gamma_n - 1/2 + \sqrt{ \Xi_n^2+ R^2 \kappa^2 }\right)\;,
\ee

Kaplan \emph{et al.} \cite{Kaplan} and Mueller  \emph{et al.} \cite{HoProgenitor,Ho} have already appreciated
the equivalence of the Schr\"odinger equation for the $1/R^2$ potential and the RG flow equations for a theory with this particular beta function. Following \cite{HoProgenitor,Ho}, we provide an interpretation of the running of the couplings:
Eq. (\ref{c4radial}) has a bad singularity at $R=0$ (a total collision, the `big bang' of the model), and therefore one will seek to regularize
the potential near the origin.\footnote{Regularizing the potential is a complementary approach to that of self-adjoint extensions because, like it, it amounts to fixing a boundary condition near the origin. But the  self-adjoint extensions approach is non perturbative and regulator-independent, so we keep both languages at hand to supplement each other and guard ourselves against artefacts of the regularization choice.} For example one could take a 
a square-well cutoff \cite{CMB}:
\be
V(R) = \left\{ \begin{array}{ll} g_n/R^2\;, ~ & R> R_0 \\ \gamma/R_0^2\;, ~�& R \le R_0\end{array} \right. \;,
\ee
Then one has to match the solution and its derivative at the cutoff, and an efficient way to do this is to match the logarithmic derivatives at $R=R_0$:
\be
\label{boundary}
\qquad \lim_{R\to R_0^+}  \frac{ R\frac{\partial\psi (R)}{\partial R}}{\psi(R)} = \lim_{R\to R_0^-}   \frac{ R\frac{\partial\psi (R)}{\partial R}}{\psi(R)} \;.
\ee
This equation implies a relation between the coupling $\gamma$ and the cutoff radius $R_0$, which causes $\gamma$ to run. If we calculate the logarithmic derivative of the wavefunction inside the cutoff radius at $R_0$, we obtain $\gamma_n (R_0)$. As a function of $R_0$, $\gamma$ satisfies Eq. \eqref{BetaFunction} \cite{Kaplan}. The radial Schr\"odinger equation is equivalent to the RG group flow equations \eqref{RGflowEquations}.

The usual interpretation in RG theory is that changing the cutoff corresponds to probing the system at different energies or, equivalently, at different scales. Actually the independent variable in an RG flow is not an energy but a ratio of energies. This matches the intuition that, if we are considering the entire universe, there is no scale external to it. 
In our cosmological model the RG flow parameter is a ratio of scales, and it gives rise to a running of the couplings $u_n$'s that redistributes the probabilities among different scale-invariant eigenfunctions. The Universe at different (relative) scales will change, as the RG flow guides the probability from one region of shape space to another.\footnote{We will not discuss here the major interpretational issue - which as yet has not definitive answer -- that every quantum theory of thel Universe must face: what is the meaning of these probabilities?} 

\subsection{Renormalization-group flows}

Let us sketch the evolution of the couplings $u_n$ under the RG flow. This determines the relative weights of the wavefunctions in \eqref{psi}. There are three kinds of behaviour:
\begin{enumerate}
\item None of the $u_n$'s corresponding to $g_n >3/4$ evolve. They are frozen at their initial values.
\item All the $u_n$'s corresponding to $-1/4 <g_n <3/4$ evolve under the RG flow between two fixed points of \eqref{BetaFunction}. One of them, which fixes $\kappa=0$, is a Friedrichs extension that effectively sets the anomaly to be zero. The other one corresponds to $\kappa=+\infty$. The $u_n$'s have the following expressions:
\be
u_n(\Rs,\theta)=\sqrt{\frac{\kappa(\theta)}{\mu} \Rs}K_{\Xi_n}\left(\frac{\kappa(\theta)}{\mu} \Rs \right) = \sqrt{\frac{\kappa(\theta)R}{\hsi}}K_{\Xi_n}\left(\frac{\kappa(\theta)R}{\hsi} \right)
\ee
Because the $u_n$'s depend only on the product $\kappa(\theta)R$ it is clear that there is a duality between the evolution under the RG flow and the evolution of $R$, from the definition $\kappa(\theta)R = R(\theta)$. The radius of the universe, measured as a dimensionless ratio relative to the initial (subtraction) point, changes under the RG flow, renormalizing the couplings $u_n$:
\be
u_n(R(\theta))= \sqrt{\frac{R(\theta)}{\hsi}}K_{\Xi_n}\left(\frac{R(\theta)}{\hsi} \right)
\ee
 The two fixed points are reached for $R=0$ and $R=\infty$. The infinite past and future are conformal: this is a realization of Strominger's proposal \cite{Strominger} that the evolution of the Universe results from the renormalization
of a conformal field theory flowing between two fixed points, associated to dark-energy dominated epochs of
accelerated expansion.\footnote{See also the model proposed recently
 in \cite{McFaddenSkenderis} and related papers.}  The structure of this model, however, is conceptually richer: we also have a frozen regime and a cosmological Efimov effect.

\item If $g_n < -1/4$ we enter the Efimov effect region. The most general expression for each $u_n$ is a linear combination of all the eigenfunctions of the extended Hamiltonian:
\be
u_n(R(\theta))= \ssum_{m} \sqrt{\kappa_m(\theta)\Rs}K_{\Xi_n}(\kappa_m(\theta)\Rs).
\ee
We have the same correspondence between evolution under RG flow and evolution of $R$ as in the weak regime. Moreover, as we saw, the limit cycle behaviour implies that the following geometric relation holds:
$\kappa_m = e^{m \pi/\Xi_n} \kappa$.

\end{enumerate}

\section{Semiclassical limit}
\label{latetime}

In this section we will study how the physics of our toy model appears at late times, far from $R=0$. In this limit we can regard $R$ as a ``heavy'' degree of freedom, in a way that we will make precise in the following discussion.\fn{We do not have the ``multiple choice problem'' of usual semiclassical approaches because any function $f$ of $R$ and $D, f(R,D),$ is singled out by the anomaly. This is a significant advantage of the shape-dynamic approach.} Without loss of generality, let us rewrite the Hamiltonian constraint \eqref{NewHamiltonianConstraint}, substituting $\psi(R,\sigma) = \eta(R)\chi(R,\sigma)$. The evolution under RG flow is not unitary, so the function $\eta$ is fixed by the requirement that the function $\chi$ stays normalized on shape space,   $\langle \chi, \chi \rangle=1$, where the scalar product is the shape Hilbert space $\ms{L}^2(S^{3N-1})$. Denoting with a prime the derivatives w.r.t. $R$,
\be
\label{WKB0}
-\hsi^2 (\eta'' \chi + 2\eta' \chi' + \chi'' \eta) - \hsi^2 \frac {3N-4}{R}(\eta' \chi + \eta \chi') + \frac{\eta}{R^2}(-\hsi^2 \Delta_{S^{3N-4}}  \chi + V_{\st{shape}}\chi) = - \kappa^2 \eta \chi \,.
\ee
Denoting by $\langle \cdot, \cdot \rangle$ the scalar product on the scale-invariant Hilbert space $\ms{L}^2(S^{3N-4})$, and assuming that $\chi$ is normalized to 1 at each value of $R$,  so that  $\langle \chi, \chi \rangle=1$, the mean over the scale-invariant (''light'') degrees of freedom of \eqref{WKB0} gives
\begin{align}
-\hsi^2 \eta'' -2\hsi^2 \eta' \langle \chi, \chi' \rangle - \hsi^2 \eta \langle \chi, \chi'' \rangle - \hsi^2 \frac{3N-4}{R} \left(\eta' + \eta \langle \chi, \chi' \rangle \right) + \\
 \frac{\eta}{R^2}[ - \hsi^2 \langle \chi, \Delta_{S^{3N-4}}  \chi \rangle + \langle \chi, V_{\st{shape}} \chi \rangle]  = -\kappa^2 \eta \, \langle \chi, \chi \rangle \equiv -\kappa^2 \eta \,. \nonumber
\end{align}
Let us suppose now that the mean values over the light degrees are vanishingly small, apart from $\langle V_{\st{shape}} \rangle=\langle \chi, V_{\st{shape}} \chi \rangle$. This is very reasonable because of our assumption that the shape degrees of freedom are very rapidly varing w.r.t. $R$.  We get
\be
\label{WKB2}
-\hsi^2 \eta'' - \hsi^2 \frac{3N-4}{R}\eta' + \frac{\langle V_{\st{shape}} \rangle}{R^2}\eta = -\kappa^2 \eta \,.
\ee
Now substitute again this equation in \eqref{WKB0} to get
\be
\label{WKB1}
-2 \hsi^2 \left(\frac{\eta'}{\eta} \right) \chi' - \hsi^2 \, \chi'' - \hsi^2 \frac{3N-4}{R} \chi' +  \frac{1}{R^2}(-\hsi^2 \Delta_{S^{3N-4}}  \chi + V_{\st{shape}}\chi -\langle V_{\st{shape}}\rangle \chi) = 0 \,.
\ee
Under the hypothesis that $\eta$ is a complex field, it can always be rewritten as $\eta(R)=\alpha(R)e^{iS(R)/\hsi}$.\footnote{The assumption that $\eta$ is complex could be unjustified because is taken to be a solution of a \emph{real} equation (see \cite{JulianComplex})}
We substitute this in \eqref{WKB2} and require the real and imaginary parts to vanish separately,
\be
\label{WKB2.5}
-\hsi^2 \, \alpha'' + \, \alpha \, S'^2 - \hsi \frac{3N-4}{R}\alpha' + \frac{\langle V_{\st{shape}} \rangle}{R^2} \, \alpha + \kappa^2 \alpha = 0
\ee
\be
\label{WKB3}
-2 \, \hsi \frac{\alpha'}{\alpha} \, S' -  \hsi  S'' -   \hsi \frac{3N-4}{R} \, S' = 0
\ee
%
Let us examine at Eq. \eqref{WKB2.5} in the classical limit defined compatibly with our previous assumptions as the case in which the amplitude $\alpha$ varies much less than the phase $S$. In this limit Eq. \eqref{WKB2.5} is a  Hamilton--Jacobi equation
\be
\label{WKB4}
 S'^2 = - \kappa^2 - \frac{\langle V_{\st{shape}} \rangle}{R^2} \,.
\ee
From $S' =  dR/dt^{\st{em}}$, the Hamilton--Jacobi equation gives us a definition of an emergent classical time; it has the dimensions of a surface,
\be
\label{WKBtime}
dt^{\st{em}}=\frac{dR}{ \kappa \sqrt{ - 1 - \frac{\langle V_{\st{shape}} \rangle}{R^2 \kappa^2}}} \, .
\ee
Let us recast physics in this emergent classical time. 
%
%
%
Substituting \eqref{WKB4} and \eqref{WKB3} in \eqref{WKB1}
\be 
-2 \, i \, \hsi S' \,\chi'  + \hsi^2  \,  \frac{S''}{S'}  \, \chi' - \hsi^2  \,   \chi'' + \frac 1 {R^2} \left( - \hsi^2 \Delta_{S^{3N-4}}  + V_{\st{shape}} -  \langle V_{\st{shape}} \rangle \right) \chi = 0
\ee
At late times we can disgregard the term $S''/S' \propto 1/R^3$.  The term in $\chi''$ measures the kinetic energy associated to the overall expansion of the Universe. If this is neglibible w.r.t. the kinetic energy associated to the all whole of the shape degrees of freedom, 
 we get the equation 
\be
\label{WKB5}
i\hsi \frac{d \chi}{dt^{\st{em}}} = -\frac{1}{R^2}(\hsi^2 \Delta_{S^{3N-4}}  \chi + (V_{\st{shape}} - \langle V_{\st{shape}} \rangle )\chi).
\ee
Because $R$ is a heavy degree of freedom, $\frac{1}{R^2} \Delta_{S^{3N-4}}  \chi \approx \Delta_{\mathbb{R}^{3N-3}} \chi$, 
 the previous equation finally becomes a time-dependent Schr\"odinger equation:
\be
\label{WKB6}
i\hsi \frac{d \chi}{dt^{\st{em}}} = -\hsi^2 \Delta_{\mathbb{R}^{3N-3}} \chi + \frac{1}{R}V_{\rm New} \chi -  \frac{\langle V_{\st{shape}} \rangle}{R^2} \chi.
\ee
This equation is very interesting because it shows what could be achieved in a more sophisticated implementation of the ideas of this toy model:
\begin{enumerate}
\item It proves that our time-dependent Schr\"odinger equation could arise (for late-time physics, in a semiclassical limit) from a completely scale-invariant cosmological model as a consequence of the scale anomaly. A cosmological force appears. It is proportional to the mean value of the scale-invariant shape potential and repulsive.
\item If we ignore the cosmic origin of the dimensionful time \eqref{WKBtime} and consider \eqref{WKB6} as the equation of an effective theory empirically determined, we would regard time as a fundamental unit and, for dimensional consistency, $\hsi$ would be given the dimensions of an energy times a time. This is why we would write \eqref{WKB6} with $\hbar$ instead of $\hsi$, where $\hbar$ is considered a dimensionful constant.
This suggests a general mechanism by which dimensionless fundamental constants in a scale-invariant model (in this case, $\hsi$) can be replaced by dimensionful constants in a semiclassical approximation. 
%
\item The fact that the `gravitational constant' depends on the `time' $R$ is a serious potential defect of the model that we do not wish to hide, especially since there are relatively strong bounds on the secular variation of Newton's constant. It should be remarked, however, that we derived Eq. \eqref{WKB6} under the assumption that $R$ is almost constant compared with the observable degrees of freedom, so we cannot use it for predicting a cosmological evolution of $G$. Moreover, we are mainly interested in this toy model as a conceptual test ground for exploring the problem of time and scale invariance. 
\end{enumerate}

\section{The Three-Body Model}


An important advantage of the three-body case is that the non-anomalous constraints ${\bf \hat P} =0$ and
${\bf  \hat L}=0$ can be explicitly reduced before quantizing. But the reason why we are studying this problem is for the conceptual insights we will get from it. First we will get a physical understanding of what the scale-invariant Planck constant means. Secondly we will propose a natural solution to the question of setting initial-value conditions in our scale-invariant model. Both issues are interesting because seems they will not be restricted just to our particle toy-model. 

\subsection{Phase-space reduction}

\begin{figure}[ht]
	\centering
		\includegraphics[width=0.6\textwidth]{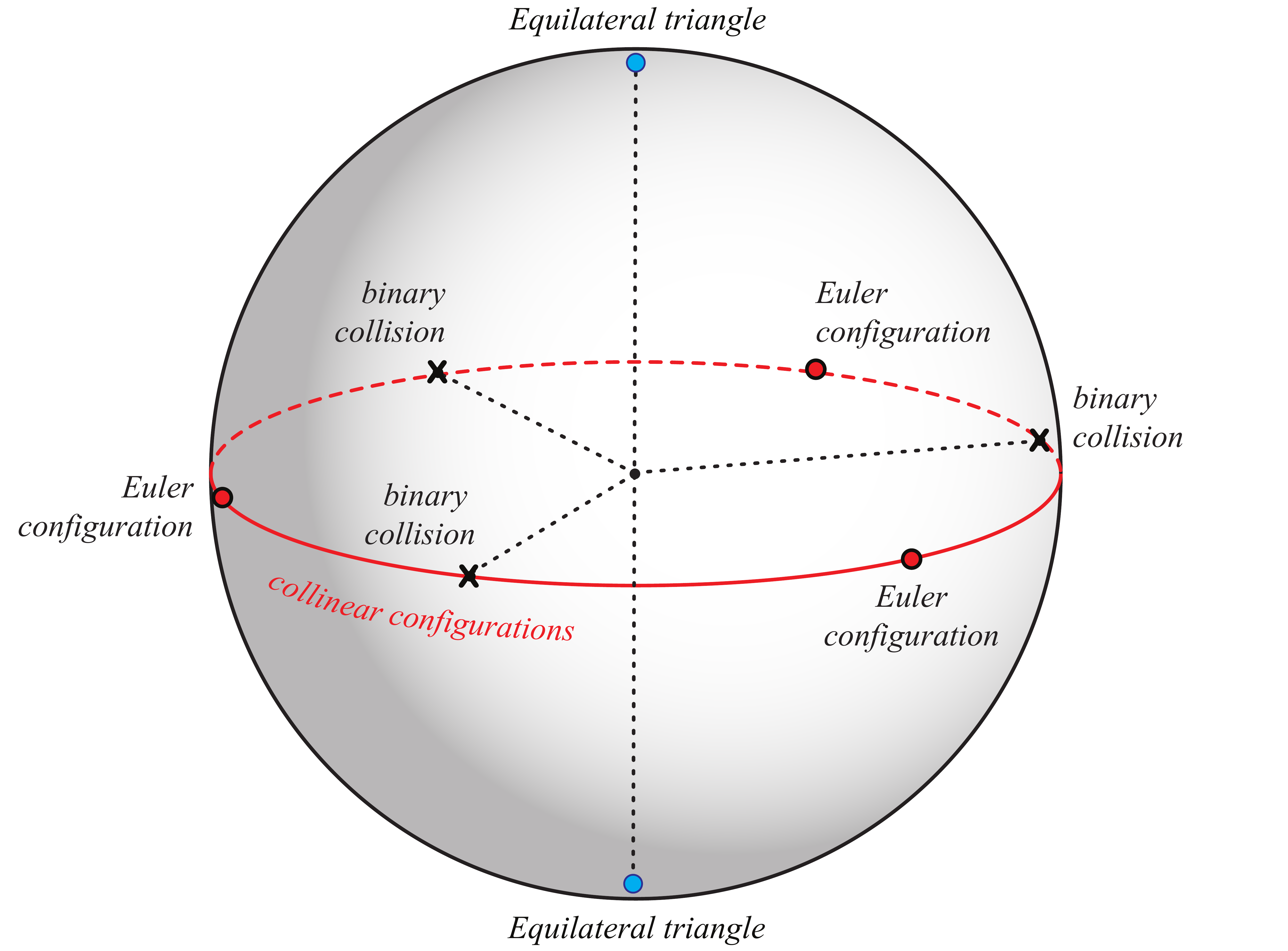}
	\caption{The shape sphere of the three-body problem. The equator corresponds to
	the collinear configurations; at three points on it there are two-body collisions, where $V_{\st{shape}}$ is singular. Between them, at positions that depend on the mass ratios of the particles, there are saddle points of $V_{\st{shape}}$ (Euler configurations). The points in the southern hemisphere represent triangles which are mirror images of the corresponding ones (with the same longitude and opposite latitude) on the northern hemisphere. The north and south poles are equilateral triangles, at which, by a result due to Lagrange, $V_{\st{shape}}$ attains its absolute magnitude for \emph{arbitrary} values of the masses. The critical points of $V_{\st{shape}}$ (Euler points and the equilateral triangles) are \emph{central configurations} and play an important role in three-body theory. For example, if the system is `held at rest' at one of them and then `released', it will fall homothetically -- without changing its shape -- to a triple collision at the centre of mass, beyond which the solution cannot be continued. The figure is for the equal-mass case, with the binary collisions evenly spaced at
	$120^\circ$ intervals.
	}
	\label{ShapeSphere}
\end{figure}

We will follow Montgomery \cite{InfinitelyManySygizes} in the phase space reduction. The first simplification comes from the fact that, because the angular momentum $\textbf{L}$ is zero, the problem is planar, so we can gauge fix
two out of three angular momentum constraints plus one translational constraint by fixing the plane in which the motion stays. We are left with three two-component vectors $\mathbf r^1$,
$\mathbf r^2$ and $\mathbf r^3$ that define the position of the three bodies on this plane.
Then we gauge fix the two remaining translation constraints by going to mass-weighted
Jacobi coordinates:
\be
\bm \rho^1 = \sqrt{ \frac{m_1 \, m_2 }{m_1 + m_2} }\left( \mathbf  r^2 - \mathbf  r^1 \right) \;, \qquad \bm \rho^2 = \sqrt{ \frac{m_3 \, (m_1+m_2) }{m_1 + m_2 + m_3} } \left(\mathbf r^3 - \frac{m_1 \, \mathbf r^1 + m_2 \, \mathbf r^2}{m_1+m_2} \right) \;.
\ee
We are left with four coordinates $\bm \rho^1$, $\bm \rho^2$ and a single angular momentum constraint to gauge fix. The coordinates
\be
w_1 = \frac 1 2 \left(||\bm \rho^1||^2 - || \bm \rho^2 ||^2 \right) \;, \qquad w_2 = \bm \rho^1 \cdot \bm \rho^2 \;, \qquad  \qquad w_3 = \bm \rho^1 \wedge \bm \rho^2 \;
\ee
are invariant under the remaining rotational symmetry and therefore give a complete coordinate
system on the reduced configuration space. The norm of the vector $\mathbf w $ is proportional  to the moment of inertia $||\mathbf w ||^2 = R^2/4$, so the angular coordinates in the three-space
$(w_1,w_2,w_3)$ coordinatize shape space, which has the topology of a sphere \cite{InfinitelyManySygizes}. This \emph{shape sphere}, as can be seen in Fig. \ref{ShapeSphere}, has several distinct regions of interest.
\subsection{The spectrum of the scale-invariant Hamiltonian}

For our purposes it suffices to say that  the scale-invariant Hamiltonian takes the form \cite{InfinitelyManySygizes}\,\footnote{Strictly speaking, the scale-invariant Hamiltonian is not
self-adjoint due to the three singularities at the two-body collisions, and a $U(3)$ family of
self-adjoint extensions would be needed. But these singularities are not as bad as the $1/R^2$ singularity at the origin, because they go like $1/R$. Moreover, these singularities 
reproduce the Newtonian gravitational interactions among particles, and there are good
reasons just to take the Friedrichs extension for all of them, which are the extensions that
allow us to best model the hydrogen atom.}
\be
\hat H_{\st{shape}} = - \hsi^2 \, \Delta_{S^2} + V_{\st{shape}} \;, 
\ee
where $\Delta_{S^2}$ is the 2-dimensional spherical Laplacian and
\be
V_{\st{shape}} = - \sum_{I<J} \frac{(m_I \, m_J)^{3/2}}{(m_I + m_J)^{1/2}(m_1+m_2+m_3)^2} \frac{1}{\sqrt{1- \cos \chi \; \cos (\psi - \psi_{IJ})}} \;,
\ee
$\chi$ is the elevation angle from the equator and $\psi$ the azimuthal angle on the shape sphere. $\psi_{IJ}$ is the azimuthal angle of the two-body collision between particles $I$ and $J$.
Moreover, our Hilbert space will be $\ms{L}^2(S^2)$, with inner product
\be
\left( f, g \right) = \int_{S^2} d \sin \chi \, d \psi \, \bar f(\chi,\psi) \, g(\chi,\psi) \;.
\ee

We will provide a qualitative description of the spectrum of $\hat H_{\st{shape}}$. We already mentioned
that it can be proved that the spectrum is discrete and bounded from below. 
Let us start from the most excited states, with positive scale-invariant energy $\hat H_{\st{shape}} \varphi_n = \lambda_n \, \varphi_n$ and $\lambda_n \gg 1$. A convenient complete orthonormal basis in which to calculate the matrix elements of $\hat H_{\st{shape}} $ is that of the eigenfunctions of the
spherical Laplacian, the spherical harmonics:
\be
- \Delta_{S^2} \, Y^m_\ell = \ell (\ell+1) \, Y^m_\ell \;.
\ee
It can be shown that in the expansion
\be
\langle Y^m_\ell  | \hat H_{\st{shape}} | Y^{m'}_{\ell'} \rangle =  \delta_{m m'} \delta_{\ell\ell'} \, \ell ( \ell+1) + \int d \sin \chi d \psi \, {\bar Y}^m_\ell \, Y^{m'}_{\ell'} \, V_{si} \;
\ee
the matrix elements of $V_{\st{shape}} $ are bounded from above, being $V_{si}$ bounded and harmonic functions normalized, while the term $\ell(\ell+1)$ grows
arbitrarily large with $\ell$. Therefore asymptotically the spectrum is well approximated by that
of the free particle on the sphere, that is
\be
\lambda_n \to \hsi^2 \, \ell (\ell+1) \;, ~~~\text{for $n$, $\ell$ sufficiently large.}
\ee
For the less excited states, one gets a hyperfine splitting from the matrix elements $\langle Y^m_\ell  | V_{\st{shape}} | Y^{m'}_{\ell'} \rangle$, which introduce an $m$-dependence of the eigenvalues.

Another regime in which we can say something qualitative about the spectrum is close to the ground state. In fact our problem is completely equivalent to that of a particle constrained on a sphere and subject to a potential that has three Coulombian wells. Close to one of the three
singularities $\chi = 0, \psi = \psi_{IJ}$, the potential is well approximated by a Coulomb potential 
and the Laplacian is approximately flat:
\be
V_{\st{shape}} \sim  \frac{\text{\textrm {const.}}}{\sqrt{\chi^2 (\psi - \psi_{IJ})^2}}  \;,\qquad \Delta_{S^2} \sim \frac{\partial^2}{\partial \chi^2} + \frac{\partial^2}{\partial \psi^2}\;, \qquad \chi,\psi ~ \text{in a neighborhood of}~�(0,\psi_{IJ}) \;.
\ee
If $\hsi$ is sufficiently small, the ground state and the first few excited states will resemble
those of the hydrogen atom (or, rather, a linear combination of those of three independent
hydrogen atoms).
The reason why $\hsi$ needs to be small enough is that $\hsi$ plays, for those wavefunctions,
the role of a Bohr radius, setting a scale of quantum phenomena \emph{in shape space}. Therefore it measures how far the wavefunction ventures from
one of the singularities at $(0,\psi_{IJ})$. If it is localized enough, it will feel neither the curvature
of the shape sphere nor the presence of the other singularities.

In the limit of small $\hsi$ we are able to calculate the ground scale-invariant energy:
it is (let us assume, for simplicity, that the three particles have the same mass)
\be
\lambda_0 \sim - \frac{1}{4 \, \hsi^2} \;.
\ee
Then from Eq. (\ref{radial}) we get 
\be
g_0 \sim - \left( \frac 1 4 + \frac  1 4 (3N-4)(3N-6) \right)\frac{1}{\hsi^2} =
-  \frac{4}{\hsi^2}
\;,
\ee
which can be made smaller than $-1/4$ if $\hsi$ is small enough.
This means that some shape eigenstates enter  the strong coupling regime.

\subsection{The problem of the initial conditions in a scale-invariant model}
\label{alphacondition}

We have, at this point, a  qualitative picture of the features of our model. At least in the three-body case,
we have established the form of  the spectrum of the scale-invariant Hamiltonian, estimated the ground-state 
eigenvalue and determined its asymptotic behaviour for large eigenvalues. This spectrum is \emph{almost} all that one
needs to determine the $R$-evolution of the wavefunction. A further bit of information is needed, in
the form of an overall normalization constant of each $u_n$ function. In fact choosing the self-adjoint
extension (\ref{NewHamiltonianConstraint}) for all $n$ only fixes the (logarithmic) derivative of $u_n$ in
Eq. (\ref{c4radial}), which we
reproduce here for convenience,
$$
- \hsi^2 \frac{\partial^2 u_n}{\partial R^2} + \frac{g_n}{R^2}u_n = - \kappa^2 \, u_n  \;,  \qquad 
g_n = \lambda_n - \frac{1}{4} (3N-4)(3N-6) \; \hsi^2\;.
$$
This is a second-order differential equation and therefore depends on two initial conditions. 
The logarithmic derivative $u'/u$ is fixed $\forall n$ by the self-adjoint extension via Eq. \eqref{c4boundarycond}, and the other can be fixed by
specifying an initial condition for the functions $u_n$, which can be set at a finite radius $R_0$,
\be
u_n (R_0) = c_n \;,  \qquad c_n \in \mathbbm{C} \;,
\ee
or at $R_0=0$ (through a limiting procedure), or even at $R_0= \infty$.\footnote{We noted earlier, in footnote \ref{capsule}, that in the case of quantization after reduction the spectrum, if it exists at all, consists solely of a zero eigenvalue. If $h_{\st{si}}$ can be `tweaked' to ensure existence of such an eigenvalue, the theory will still be incomplete if the eigenvalue is degenerate and corresponds to a superposition of zero-eigenvalue wavefunctions.}

We propose now a way of fixing this initial condition; it relies on the \emph{highly asymmetric} structure of
shape space.  
In fact, $C=-V_{\st{shape}}$, which may be called the \emph{complexity}, defines a positive-definite function on shape space that takes its minimum value on the configuration of the system that is more uniform than any other configuration. In the case of the three- and four-body problems for \emph{arbitrary} values of the masses \cite{BattyeGibbons}, the minimum of $C$ is at the equilateral triangle and regular tetrahedron respectively.
Battye et al have done numerical calculations for up to $10^4$ particles which showed that the minimum of $C$ is realized on remarkably uniform (super-Poissonian) configurations \cite{BattyeGibbons}.

The fact that all relative configuration spaces have a distinguished point was noted by one of us many years ago \cite{Barbour_Niall:first_cspv,barbour:timelessness,barbour:timelessness2} but before the importance of quotienting wrt dilatations as well as translations and rotations was fully appreciated. The distinguished point identified in \cite{Barbour_Niall:first_cspv,barbour:timelessness,barbour:timelessness2} corresponds to the configuration with all particles collapsed to a single point and, for obvious reasons, was called Alpha. The additional quotienting wrt dilatations and the associated passage to shape space makes it possible to transform the qualitative intuition of the original idea into a quantitative theory because $C$ is an objective \emph{scale-invariant} measure of complexity on shape space. It is a classical and quantum-mechanical \emph{observable}. Simultaneously -- and suggestively -- it is, as we have shown, (minus) the potential that governs the scale-invariant dynamics.


We now define Alpha to be the most uniform shape of the system, where $C$ has its minimum.\footnote{There is no corresponding notion of a most complex state:
there is an Alpha, but no Omega \cite{barbour:eot,barbour:scale_inv_particles,JulianComplexityPaper}.} 
The presence of the Alpha point in shape space suggests the existence of a distinguished \emph{initial condition} for the wavefunction of the universe: at early times it could be peaked around Alpha.
When this choice is made, the expectation value of $V_{\st{shape}}$ is close to its maximum, the complexity close to a minimum. 

\section{Discussion and Conclusions}

In this paper we have discussed the so-called problem of time in quantum gravity using an analogue particle model whose algebra of constraints is similar to the one of GR in Hamiltonian form. This model is 
almost uniquely fixed by relational first principles identical to those that one of us proved to be
at the basis of GR \cite{barbourbertotti:mach,JuliansReview}.

As we mentioned in the introduction, our basic idea is that two problems in GR could end up providing the solution to each other. We explored this possibility in the particle model. The two problems are the frozen nature of the Wheeler--DeWitt equation and the failure of GR in its usual representation\footnote{The caveat ``in its 
usual representation'' relates to the possibility noted in footnote \ref{FootOnTheOtherTwoPapers} of 
achieving scale invariance by trading a global degree of freedom for an 
internal time.} to be fully relational: a single global degree of freedom, the volume of the Universe, cannot be considered as gauge.  
In this paper we insisted on strict classical relational principles by choosing a fully \textit{scale-invariant} analogue model. We then discovered that time and scale emerge naturally through an anomaly that arises if we attempt a consistent Dirac canonical quantization.

We concluded that in Dirac quantization of the analogue model global scale invariance is an anomalous symmetry that is broken in the quantum theory. This is due to the singularity that the potential shows at a total collision. Its renormalization leads to an RG flow in which the radial eigenfunctions $u_n$ in the expansion
\be
\psi(R,\sigma) = \sum_n \,  \, u_n(R) R^{(4-3N)/2} \, \varphi_n (\sigma) \;, \qquad \sigma \in S^{3N-4} \;,
\ee
behave like RG running coupling constants for the shape eigenfunctions $\varphi_n$.
In particular their logarithmic derivative satisfies a Callan--Symanzik equation. The behaviour of the couplings $u_n$  in Eq. \eqref{RGflowEquations} depends on the scale-invariant energy $g_n$ associated to $\varphi_n$ (see Eq. \eqref{scaleinvariantproblem} and \eqref{radial} for the definition of $g_n$). Above a certain limit, the couplings are frozen; then there is a regime in which they flow between two conformal fixed points; and finally the lowest energies exhibit an Efimov behaviour with limit cycles. 

In this intriguing picture the  
dimensionless ratio of the scale of the universe relative to a reference value becomes `time' in the sense that the changing scale of the universe renomalizes the couplings $u_n$, and this corresponds to a flux of probability on shape space. The scale ratio works this way because its change produces an RG flow that redistributes the relative probabilities among the shape eigenfunctions. We call this proposal \emph{double emergence}, because scale and time emerge from the same quantum phenomenon, an anomalous symmetry breaking.\fn{Admittedly this model is still affected by the quantum-mechanical measurement problem. The wavefunction in fact spreads over macroscopically distinguishable configurations in shape space.}

What we have produced is a concrete implementation of the holographic cosmology scenario (albeit for a particle toy-model) that has very sound principles attached to it. In fact, our motivations have nothing to do with holography at all. This could be interesting from the point of view of finding a physical foundation of the AdS/CFT correspondence.

Besides this main result, in this paper we obtained a number of other interesting results
and observations, which we summarize here.
\begin{enumerate}
\item {\bf The semiclassical regime.} We studied the wavefunction of the system
at large values of our `time' $R$, under the assumption that the wavefunction
fluctuates much more rapidly in the shape directions than in $R$ (Born--Oppenheimer
approximation). With a WKB ansatz, we were able to find an effective time-dependent
Schr\"odinger equation satisfied by the wavefunction, with a function of $R$ and
$\langle V_{\st{shape}} \rangle$ playing the role of time.
This effective semiclassical dynamics is expected to become dominant at late
times (large $R$'s). Our equation contains a repulsive cosmological force and a dynamically fixed Newton constant $G=1/R$, but
the equation is not valid if $R$ varies at cosmological scales.


\item {\bf The dimensionless Planck constant.} The `Planck constant' that appears in the
canonical commutation relations of our model is a dimensionless number. This is due
to the fact that the action is dimensionless, and the momenta have dimensions of 
inverse lengths. In the semiclassical, `late-time' regime we were able to show
a mechanism in which the familiar, dimensionful Planck constant could emerge.
At the fundamental level time has no meaning, and the quantity that emerges at
the classical level as time has the dimensions of a length squared. If one attributes
to it a fundamental role, as is done when dealing with the effective physics that
is relevant in the laboratory, then the Planck constant takes on some dimensions,
which are needed to convert from squared lengths to times.
This role of the Planck constant is identical to that of Boltzmann's constant in thermodynamics:
fundamentally it is dimensionless, but at the effective level one attributes to it the
dimensions that are necessary to convert from energy to heat, which at the fundamental
(in that case, microscopical) level are the same thing.
Even if eventually the particular toy model considered in this paper gets ruled out,
we expect such a role for the Planck constant to emerge whenever one assumes
that the fundamental degrees of freedom are dimensionless.

\item {\bf The three-body problem.} We studied the case of three bodies, which is
unique for the important simplifications that it allows: the angular momentum constraint
can be algebraically solved, and shape space can be conveniently described as a two-dimensional
sphere, whose radius is proportional to the moment of inertia $R$, our `time'.
The shape Hamiltonian is simply the spherical Laplacian on the shape sphere
plus $V_{\st{shape}}$, which is a relatively simple function on the sphere.
A qualitative study of its spectrum was enough to establish that all of the three regimes of the RG flow are realized. Moreover the scale invariant Planck constant appeared as a scale of quantum phenomena on shape space.

\item {\bf The initial condition.} The only thing that this model leaves unspecified
is the initial condition for the wavefunction on shape space, which can be fixed by
setting the coefficients $u_n$ at a certain value of $R$. Developing the
classical proposal of a distinguished `$\alpha$-point' in the configuration space of the scale-invariant degrees of freedom made in 
\cite{barbour:eot}, we suggest that the driving potential of the scale-invariant dynamics,  $V_{\st{shape}}$, is also a natural 
measure of complexity in the system. In fact its maximum is attained the most uniform state, which we called the `$\alpha$-state' . \footnote{See \ref{alphacondition} for a discussion which generalize this proposal to N body problems.} So a natural initial condition for the wavefunction
on shape space is such that all the probability is concentrated on this uniform shape, the equilateral
triangle in the three-body problem. The evolution would then make
the wavefunction spread out of that point, driving change in shape.

\end{enumerate}

An interesting possibility is to extend these ideas to more general contexts. In particular it is intriguing to ask if GR could be considered as a fully conformal gauge theory on conformal superspace (the geometrodynamic counterpart of our shape space) in which a single global gauge degree of freedom becomes dynamical and a preferred notion of cosmological time emerges. The theory of shape dynamics \cite{JuliansReview,gryb:shape_dyn,Gomes:linking_paper,Barbour:new_cspv,barbour_el_al:physical_dof} could be the natural context in which these ideas can be explored.

\section{Appendices}

\subsection{Jacobi coordinates}
\label{JacobiCoordAppendix}

A simple algorithm \cite{LimBinaryTree} for creating Jacobi coordinates requires specification of a complete directed binary tree, whose leaves represent the $N$ particles, and the $N-1$ internal vertices are associated
with the Jacobi coordinates. The algorithm works this way: given such an arbitrary
tree, one assigns to each ``parent'' vertex four numbers: the three coordinates of the centre of
mass of the two ``child'' vertices and the sum of their masses, which are taken of course
from the leaves.

\begin{figure}[h!]
\includegraphics[width=0.8\textwidth]{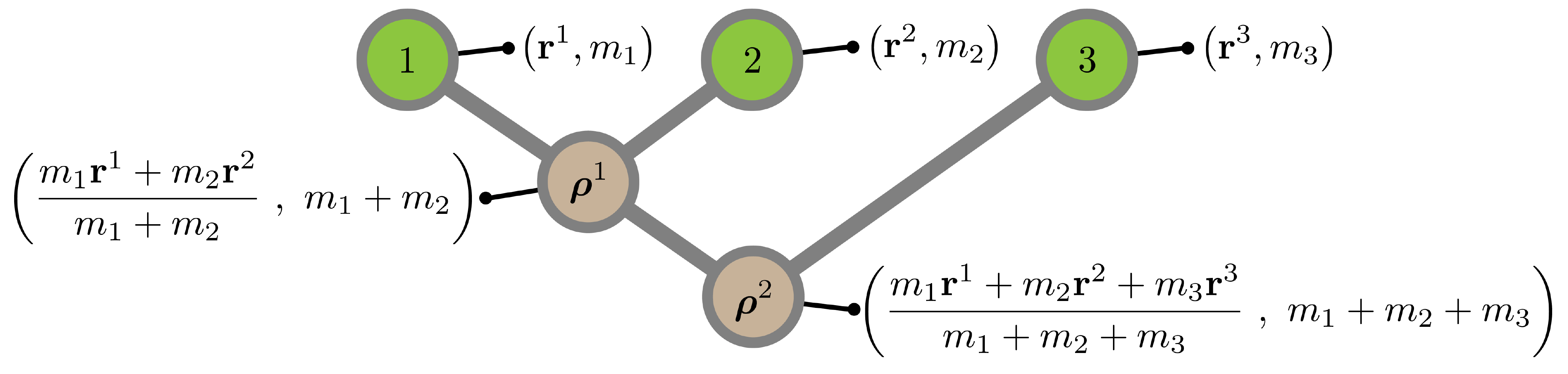}
\caption{The binary tree algorithm for the three-body system: the two Jacobi coordinates associated with the
internal nodes are ${\bm \rho}^1 ={\bf r}^1 -{\bf r}^2$, ${\bm \rho}^2 = (m_1 {\bf r}^1  + m_2 {\bf r}^2)/(m_1+m_2)  -  (m_1 {\bf r}^1 + m_2 {\bf r}^2 +m_3 {\bf r}^3)/(m_1+m_2+m_3)$.}
\end{figure}

Then one defines the Jacobi coordinates as a function on the 
internal nodes. This function associates to each node the
difference of the coordinates of its two ``child'' nodes: ${\bm \rho}^\text{parent} ={\bf r}^\text{left} -{\bf r}^\text{right}$. When applied to the root node, this function
gives the $N$-th Jacobi coordinates, that is, the coordinates of the 
centre of mass of the system, which has now been decoupled
and can be discarded, so that one can work with just with the first $3N-3$ coordinates. 
The great advantage of the coordinates, besides the centre-of-mass decoupling, is that they leave the kinetic metric diagonal \cite{LimBinaryTree}.
Its eigenvalues in this coordinate system are the Jacobi
effective masses $\mu_{\st{\,J}}$, which are just the masses of the nodes
associated with each Jacobi coordinate.
This algorithm represents a relational way to move to the centre-of-mass reference frame.

\subsection{Diagonalization of the inertia tensor}

We can solve the angular momentum constraint with a gauge-fixing,
that is, with a choice of orientation of our axes. A global gauge-fixing, though,
has to satisfy certain requirements to be good. The main requirement is that
the constraint surface defined by the gauge fixing must never be 
parallel to the vector field generated by the constraint it is supposed
to gauge fix. This translates into the requirement that the Poisson
bracket between the gauge fixing and the constraint must be
invertible, which means it must be everywhere (weakly) non-vanishing.

A popular choice of axes among $N$-body specialists is the one defined by the principal
axes of the moment-of-inertia tensor \cite{LittlejohnReinsch}. It is attractive for its simplicity and symmetry,
and more than anything else for its relational character, but it fails
to be global for the reason mentioned above: it becomes degenerate 
at certain points of configuration space.

The moment-of-inertia tensor is defined as
\begin{equation}
I^{ab} =   \sum_{I=1}^N m_I \, (r^I_a - r^\text{cm}_a) (r^I_b - r^\text{cm}_b) - \delta^{ab}  \sum_{I=1}^N m_I  \, || {\bf r}^I - {\bf r}^\text{cm} ||^2  \,.  
\end{equation}

Its eigenvectors are the three principal axes of inertia of the $N$-body configuration.
The matrix $I^{ab}$ is symmetric, and therefore can be diagonalized with a rotation
that puts the three principal axes of inertia respectively on the $x,y$ and $z$ axes. 
The three constraints that impose such a condition are the ones that set the off-diagonal
elements to zero:
\begin{equation}
 I^{12} = I^{23} = I^{13} = 0  \,, \label{diagonalize}
\end{equation}

This condition, though, does not always uniquely fix the axes for two reasons:

\begin{enumerate}
\item
When the system is in a symmetric configuration
(spherically or axisymmetric) there are identical eigenvalues (two if axisymmetric,
three if spherically symmetric), and the matrix is already (block) diagonal.

\item
Even in the non-symmetric case, the solution to equation (\ref{diagonalize})
is not unique: there are three solutions, corresponding to the residual symmetry
under exchange of the axes. But this is harmless as one can resolve the ambiguity by requiring the eigenvalues to be arranged in order of magnitude: $I^{11} < I^{22}  <  I^{33}$.

\end{enumerate}

\section*{Acknowledgements}
We would like to thank S. Gryb for his initial input that proved crucial for the beginning of this project. 
We thank also J. Louko, P. Hoehn and G. Canevari for useful comments and discussions
during the preparation of this paper.  M.L. thanks St. Hugh's College for hospitality when working on his Master Thesis in a joint exchange programme with Collegio Ghislieri; he also thanks the Institute for Advanced Studies of Pavia for partial funding.

This work was supported by  a grant from the Foundational Questions Institute (FQXi) Fund, a donor advised fund of the Silicon Valley Community Foundation on the basis of proposal FQXi 
Time and Foundations 2010 to the Foundational Questions Institute.
It was also  made possible in part through the support of a grant from the John Templeton Foundation. The opinions expressed in this publication are those of the author and do not necessarily reflect the views of the John Templeton Foundation.
Research at Perimeter Institute is supported by the Government of Canada through Industry Canada and by the Province of Ontario through the Ministry of Economic Development and Innovation.

\bibliographystyle{utphys}
\bibliography{bibanomaly}
\end{document}